# Anomalous electrowetting of physicochemically heterogeneous surfaces

Rumal Singh, Donjo George, Prashant Hitaishi[#], Samarendra P Singh, Sajal K Ghosh[*]

*Department of Physics, School of Natural Sciences, Shiv Nadar Institution of Eminence,*

*NH-91, Tehsil Dadri, Gautam Buddha Nagar, Uttar Pradesh-201314, India*

*\*Email: sajal.ghosh@snu.edu.in*

## Abstract

In the present work, a physiochemically heterogeneous surface has been fabricated to investigate the electrowetting behaviour of the surface. The polystyrene (PS) micro-humps with varied size are developed on the polydimethylsiloxane (PDMS) surface, which show an anomalous electrowetting behaviour. The surfaces are observed to be more electro-wettable than it is predicted by the classical Lippmann-Young equation. The observations are well understood considering the chemical heterogeneity of the surface, exhibiting a surface energy mismatch between the PS micro-humps and the PDMS layer. Further, the anomaly is comprehended by following the ridge formation around the triple-phase contact line and the varied surface roughness. A surface parameter is introduced in the Lippmann-Young equation that follows the experimental data with varied values of the parameter representing the physicochemically heterogeneous surfaces. A positive value of the surface parameter indicates strong pinning while a negative value represents depinning of the droplet. This parameter explains the faster electrowetting than predicted by the Lippmann-Young equation.

[#]*Present address: Institute of Experimental and Applied Physics, Kiel University, Germany*

# 1. Introduction

Electrowetting is a phenomenon where the wetting property of a surface is enhanced by the application of an external electric field. In this process, if the liquid is in direct contact with a metal surface, it limits the applications of the surface. It also brings the risk of electrolysis, which can degrade a device [1]. Inserting a dielectric layer between the metal surface and the liquid droplet makes it safe and reliable, which prevents the electrolysis and degradation. This field of research is termed 'electrowetting on dielectric (EWOD)', which opens a wide range of practical and emerging applications, especially where precise and programmable droplet manipulation is needed [2,3]. Recent applications of EWOD include transporting liquids to create optical switches [4], cooling electronic circuits with cold drops [5], mixing micro-drops for printing [6], suctioning liquids into microtubes [7], developing variable focal lenses [8] and generating electronic paper [9]. The analysis of chemical composition of liquids, like blood and urine, using lab-on-a-chip technology became superior by utilising this EWOD [10].

In an EWOD experiment setup, an electric field is applied across the dielectric layer, which is placed between a conductive surface and a liquid droplet. The system stores electrostatic energy, which arises due to the applied field, leading to enhance in energy of the interface. Such a high energy surface facilitates a water droplet to spread on the surface. This wetting of the surface as a consequence of the applied field can be quantified by measuring the contact angle expressed by classical Lippmann-Young (L-Y) equation [11],

$$cos\theta_E = cos\theta_0 + \frac{1}{2}\left(\frac{C}{A}\right)\frac{V^2}{\gamma_{lg}} = cos\theta_0 + \eta \qquad (1)$$

where, $\theta_E$ is the contact angle after applying an external voltage $V$, $\theta_0$ is the contact angle without applied voltage, $C$ is the capacitance measured over the surface area $A$ and $\gamma_{lg}$ is the liquid-gas interfacial tension. The term, $\eta = \frac{1}{2}\left(\frac{C}{A}\right)\frac{V^2}{\gamma_{lg}}$, is known as 'electrowetting number'.

This equation is valid for an ideal situation that ignores the contact angle saturation, triple phase contact line pinning, softness of the dielectric layer, electrochemical reactions and the chemical heterogeneity of the surface [12, 13]. It also neglects the roughness of the dielectric surface [13]. This roughness is important as it dominantly alters the wetting nature of a surface, converting a hydrophilic surface to a more hydrophilic and a hydrophobic surface to a superhydrophobic one [14]. If a hydrophobic surface with physical roughness shows wetting in Wenzel state, its electrowetting behaviour can be analysed by introducing a roughness factor ($R$) to L-Y equation as [15],

$$cos\theta_E = R(cos\theta_0 + \eta) \qquad (2)$$

If the surface texture or morphology signifies the Cassie-Baxter state, then the electrowetting phenomenon follows the expression,

$$cos\theta_E = f_1(cos\theta_0 + \eta) - (1 - f_1) \qquad (3)$$

where $f_1$ is the solid fraction in contact with liquid, consequently, the fraction in contact with air is $(1 - f_1)$. In these two cases, the surface is assumed to be fabricated from a single type of material on which the physical roughness exists. Therefore, though they are physically heterogeneous, they are chemically homogeneous. However, most of the natural surfaces are chemically heterogeneous with a hierarchical roughness [16]. Such a complex surface could be engineered, generating a primary surface of a material on top of which another material can be deposited with spatial heterogeneity. The deposited secondary structures can produce a macroscopic roughness on the primary surface, along with their own inherent microscopic roughness, leading to a chemically and structurally heterogeneous surface with hierarchical roughness. A systematic exploration of such an engineered complex surface is rare.

Polystyrene (PS) is a synthetic aromatic hydrocarbon polymer made from the monomer styrene. It is brittle, hard, rigid, and thermally stable up to ~200°C, which is hydrophobic in nature [17]. Polydimethylsiloxane (PDMS) is a silicone-based polymer characterized by its flexible siloxane backbone and methyl side groups, which give it exceptional elasticity, chemical inertness, and biocompatibility [18]. This hydrophobic polymer is frequently utilized in biomedical devices, microfluidics, and soft lithography [19]. In this present study, first, PDMS was spin coated on an indium tin oxide (ITO) coated glass substrate to fabricate a thin polymeric film. Then, PS solution was drop-casted on the film. Because of surface energy mismatch, micro humps of PS were generated on the PDMS surface with identical shape but varied range of size. The electrowetting behaviour of this complex surface, which is physically and chemically heterogeneous, has been investigated to ascertain if it follows the classical L-Y equation.

## 2. Materials and methods

### 2.1 Materials

Polydimethylsiloxane (PDMS), polystyrene (PS) (molecular weight 35000 g/mol), ethanol, chloroform, and sodium chloride (NaCl) were purchased from Sigma-Aldrich which were used without further purification. For the PDMS, oligomer and the crosslinker were mixed in 10:1. The indium tin oxide (ITO) coated glass substrate was purchased from Techinstro (India) while the platinum wire (diameter-100 μm, purity~99.95 %) was obtained from Manilal Manglal &

Company (India). Mili-Q, Millipore water with a resistivity of ~18 MΩcm and pH ~ 6.9 was used in preparing all the samples. A 2450 Source Meter (Keithley, Tektronix) was used to supply the voltage as well as to measure the voltage and current in electrowetting circuit.

**2.2 Substrate preparation**

To prepare substrates, the ITO-coated glass was cleaned with alternate 4 cycles of bath sonication with water and isopropanol. Later, the substrates were dried with a gentle flow of $N_2$ and placed in a UV-$O_3$ environment for 30 minutes at 100 °C to remove any organic contaminants. For PDMS film, the monomer and crosslinker were mixed in hexane, which was spin-coated on the cleaned ITO-coated glass substrates in two steps: first, at 500 rpm for 30 seconds and then at 5000 rpm for 60 seconds. Then the substrates were placed in a hot air oven (Ambinova Technologies, India) at 95°C for 24 hours for annealing. The coating of the dielectric layer of the PDMS provides the hydrophobicity to the substrate. Excess monomer or the crosslinker on the surface was removed by dipping the substrates in ethanol for 24 hours at room temperature. The final coated dielectric substrates were drop casted with varying concentrations of chloroform solution of polystyrene (PS) at room temperature with a relative humidity (RH) of 45±1 %. When the solvent was evaporated completely, the substrates were placed over a hot plate at 80° C for 30 minutes to settle the PS on the PDMS surface. These substrates were used for electrowetting measurements as schematically shown in Fig. S1 in the *Supporting Information (SI)*.

**2.3 Contact angle measurement and electrowetting method**

An optical tensiometer (One Attention Theta Flex, Biolin Scientific, Sweden) was used to monitor the in-situ contact angle during the electrowetting. For this study, a drop of 4 μL of water was placed on the substrate, and the contact angle, during electrowetting was monitored and quantified using the Young-Laplace method [20]. To complete the electrowetting circuit, one electrical connection was made with the ITO-coated substrate, and another with a platinum wire dipped in a droplet of 1 mM NaCl aqueous solution placed on the substrate, as illustrated in Fig. S1 in SI.

**2.4 Field emission scanning electron microscopy (FESEM)**

To visualize the three-dimensional structure formed by polystyrene on the PDMS film, the field emission scanning electron microscopy (JSM-7610FPlus, JEOL) was used. The substrates were gold-coated to make a protective barrier that reduces direct electron beam interaction. It also improves signal quality. The top as well as cross-sectional views of the substrate were captured.

### 2.5 Atomic force microscopy (AFM)

The surface morphology was examined with an atomic force microscope (AFM) (MFP-3D Origin). The images were generated by the non-contact mode as the substrates were coated with a soft material. The images were analysed with Gwyddion software.

### 2.6 Capacitance measurement

The capacitance of the composite layer of PDMS and the PS micro-humps was measured by an LCR meter (ZM2376, NF Corporation). For electrical contacts, multiple copper pads were deposited by thermal evaporation (BC-300, HHV, India) through a dotted mask (diameter ~ 500 μm). Capacitance-frequency and phase-frequency measurements were recorded using a 100 mV AC excitation over the frequency range 10 Hz - 1 MHz. The capacitance value was taken at frequencies where the phase angle was approximately 90º. For each substrate, measurements were performed at ten different copper contacts, and the reported value represents the statistical average of these measurements.

## 3. Results and Discussion

### 3.1. Fabricating micro-humps on a hydrophobic surface

Although it is well reported that PS and PDMS individually form a thin layer on a substrate [21,22], the aim of present work was to fabricate a distribution of well-defined microdots of PS on a dielectric layer of PDMS to fabricate a physicochemically heterogeneous surface. Upon evaporation of chloroform from the drop-casted PS solution on the PDMS layer, the PS molecules self-assemble to form such dots, as shown in Fig. 1. The top views of the FESEM images of the dots are shown in Fig. S2. The cross-sectional FESEM images confirm them to form a spherical cap or a hump on the surface (Fig. 1 (b-e)). The base radius ($R$) and the height ($h$) of the hump depend on the PS concentration. However, beyond a critical concentration of ~80 mg/ml in chloroform, the size of the humps becomes large enough to connect each other, resulting in a thin layer of PS on top of the PDMS layer. As described in the inset of Fig. 1 (*f-i*), the peak position of the Gaussian distribution of hump diameters shifts from 2.36 to 5.98 μm as the PS concentration increases from 0.5 to 10.0 mg/ml. It describes that the characteristic hump size, given by the diameter ($2R$), increases with the increasing PS concentration. Such a systematic shift of the centre of the distribution is accompanied by an enhanced standard deviation evident in the broad distribution (Table S1 in section S2 of SI). As shown in Fig. S3, the PDMS surface is more hydrophobic compared to a pure PS surface, as PDMS has a low surface energy, typically in the range of 19-21 mJ/m² while PS has a higher surface energy of 40-44 mJ/m² [23]. The contact angle hysteresis (CAH) is greater for the PDMS surface compared to the PS surface (Fig. S3), indicating the PDMS surface to be rougher one. The direct AFM imaging of these surfaces, as shown in Fig. S4, confirms this hypothesis. Due to the surface energy mismatch, PS does not like to spread on the PDMS surface. Therefore, instead of forming a smooth thin film, PS prefers to minimize its contact area with PDMS. During the evaporation of chloroform, the increasing concentration of PS initiates the triple phase contact line (TPCL) to recede. It causes a fluid flow directed inward, pulling the PS with it. It leads to PS accumulation in the centre rather than at the edge. As a result, PS assemble into small blobs forming the micro-humps. The presence of PDMS between the PS blobs ensures the surface to be chemically heterogeneous.

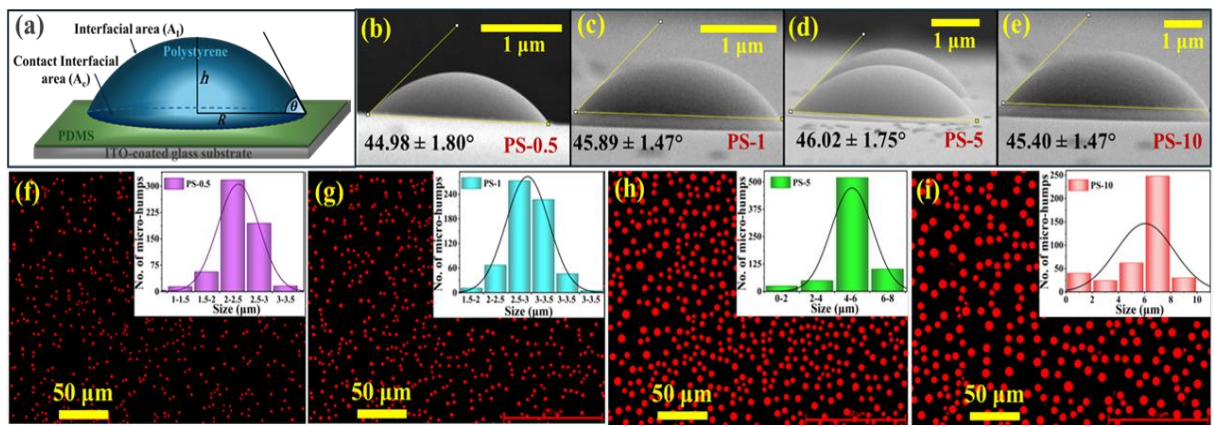

*Figure 1: (a) Schematic diagram of a polystyrene (PS) micro-hump on polydimethylsiloxane (PDMS) surface showing its base radius R and height h. FESEM images of the cross-sectional views of the humps formed by varying concentrations of PS as (b) 0.5, (c) 1.0, (d) 5.0, and (e) 10.0 mg/ml. The corresponding static contact angle (CA) is noted on the respective figure. (f – i) Optical top-view images of the humps distributed on the PDMS surface with the respective concentrations mentioned in (b – e). Corresponding size distributions are shown in the inset of the figures, with the solid curve showing a Gaussian distribution. Diminishing size of the scale bars in figures (b – e) exhibits the size of the humps to increase with increasing concentration of the PS, which is also depicted in the shift of the peak of the Gaussian distribution*

On the unit surface area of PDMS, the surface coverage by PS micro-humps increases with increasing PS concentration (Fig. 2(a)). As a result, the net surface energy increases, which is quantified in section S5 and shown in Fig. S5. These micro-humps increase the surface roughness slightly, as shown in Fig. 2(b), which is quantified in section S6. The advancing contact angle ($\theta_a$) increases, while the receding contact angle ($\theta_r$) decreases with increasing PS concentration (Fig. 2(c)). Thus, the contact angle hysteresis (CAH) which is the difference of

$\theta_a$ and $\theta_r$, increases significantly from the value of 30° to 50° from no PS humps on PDMS surface to the PS humps formed at a concentration of 10 mg/ml. For a hydrophobic surface, when the roughness is significantly enhanced without changing its chemical nature, the $\theta_a$ rises while $\theta_r$ drops resulting into higher CAH [14,24]. On changing the surface chemistry keeping the physical roughness constant, both $\theta_a$ and $\theta_r$ increases for the hydrophobic surface [25,26]. In the present work, the formation of isolated PS humps on PDMS surface makes the surface unique with modified roughness and surface chemistry. The boundary of PS micro-humps that meets PDMS surface and the mismatch in surface energy of PS and the PDMS cause a Gibbs surface barrier [27]. It becomes the source of pinning and depinning (stick and slip) of triple phase contact line (TPCL) of a water droplet during its advancement and recession, respectively. When the TPCL advances on the surface, it sticks increasing the $\theta_a$ slightly, while on receding, it pins strongly resulting in decreasing $\theta_r$. Thus, the difference between $\theta_a$ and $\theta_r$ increases with PS concentration (Fig. 2(c)). It is worth mentioning that the observation here is due to the physical roughness and chemical heterogeneity though the effect of the roughness is weak as the roughness factor changes slightly with enhanced number of the PS humps [28].

s

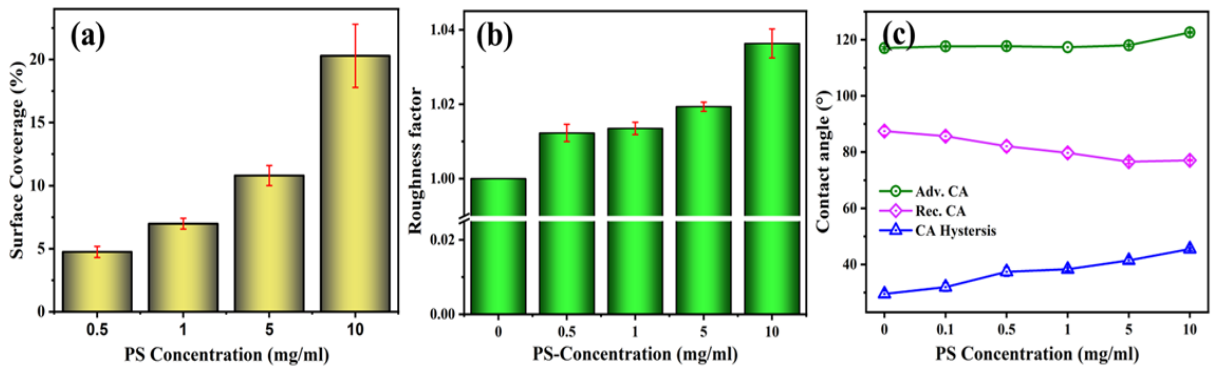

*Figure 2*: *(a) Surface coverage by polystyrene (PS) micro-humps on a polydimethylsiloxane (PDMS) surface. (b) Overall roughness factor of the PS hump-coated PDMS surface. (c) wetting nature of the surface with advancing (Adv.) and receding (Rec.) contact angle (CA) along with contact angle hysteresis. The surface parameters are plotted as a function of PS concentration. With the increasing concentration, the surface coverage increases with an enhanced contact angle hysteresis.*

### 3.2 Specific capacitance of the composite surface

From the electrowetting perspective, it is critical to understand how the addition of PS micro-humps affects the specific capacitance of the PDMS layer. The capacitance, $C = \frac{K\epsilon_0 A}{d}$, of a parallel plate capacitor with dielectric constant $K$ and thickness $d$, is directly proportional to its surface area $A$. The presence of micro-humps increases the actual surface area by 1.23, 1.35, 1.93, and 3.63 % at the PS concentrations of 0.5, 1.0, 5.0, and 10.0 mg/ml, respectively. Consequently, the capacitance is expected to rise. However, as shown in Fig. 3(a), the measured specific capacitance of PDMS film decreases in the presence of PS micro-humps. Therefore, it is interesting to explore this ambiguity by systematically investigating the shape and size of these micro-humps. Their base diameter, *2a*, varies from 2 to 8 μm, maintaining a similar contact angle of ~45° (Fig. 1(b-e)). Their height varies from 0.4 to 2 μm. The thickness of the PDMS film is measured to be 4.92 ± 0.54 μm (Fig. S7). Therefore, the effective capacitance of the composite surface reaches a value predominantly controlled by the height of the micro-humps rather than the surface area. The dielectric constant of PS with a value of 2.5 [29], which is slightly less than that of PDMS with a value of 2.66 [30], also favours to drop the capacitance.

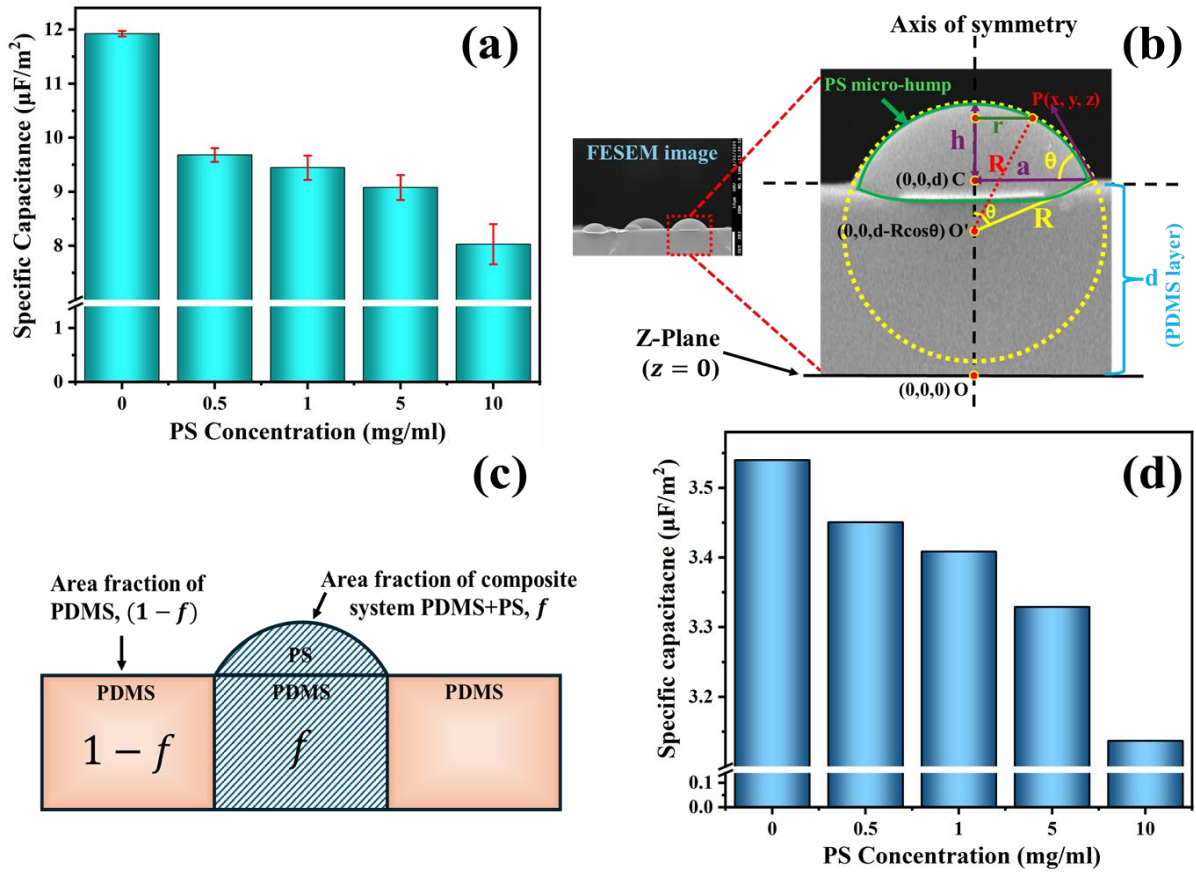

*Figure 3*: (a)Measured specific capacitance of composite surface. (b) FESEM image with fabricated PS humps on PDMS surface with a base radius of 'a' and height h. The hump could be considered as a section of a sphere of radius 'R'. (c) Schematic representation of area fraction of PDMS and composite system to calculate the effective capacitance of the system. (d) Theoretically calculated specific capacitance with increasing PS concentration.

The capacitance of the composite surface with a flat layer of PDMS covered with the micro humps of PS can be calculated by considering the geometry of the surface. Following the image shown in Fig. 3(b), each hump is considered as a cap of a sphere of radius $R$ and $h$ is the height of the hump from its base. As derived in section S8 of SI, the specific capacitance of the hump of the PS and the PDMS layer below the hump (shaded region in Fig. 3(c)) can be expressed as,

$$C_c = -\frac{2K\varepsilon_0}{a\sin\theta}\ln\left\{1 - \frac{a\sin\theta}{2(d+h)}\right\} \quad (7)$$

As $\frac{a\sin\theta}{2(d+h)} < 1$ for this physical system, $\ln\left\{1 - \frac{a\sin\theta}{2(d+h)}\right\} < 0$ which leads to the value of $C_c$ to be always positive. The effective specific capacitance ($C_{Eff.}$) of the system composed of the PDMS layer ($C_l$) (unshaded in Fig 3(c)) and the composite region (shaded in Fig 3(c)) can be calculated from,

$$C_{Eff.} = f.C_c + (1-f).C_l \quad (8)$$

where $f$ is the area fraction of micro-hump and $(1-f)$ is the area fraction of the pure PDMS layer. The calculated results from Eq. (8) are shown in Fig. 3(d) which show a systematic drop of capacitance as a function of the PS concentration. This trend closely follows the experimental observation shown in Fig. 3(a). Although the experimental values follow the same trend and order of magnitude as the calculated ones, discrepancies in the absolute values may arise as the contribution of microscopic roughness of the humps alters the effective surface area compared to that assumed in the model. Furthermore, for simplicity the capacitance expression assumes identical relative permittivity ($\varepsilon_r$) or dielectric constant ($K$) for both PDMS and PS, which is not correct, since their dielectric constants differ. Finally, the value of $K$ considered to be 2.6 in the calculation may differ from that of the real samples due to impurities in the system.

### 3.3 Electrowetting of micro-hump coated surface

Before understanding the nature of electrowetting of the composite heterogeneous surface, the behaviour of pure PDMS and PS surfaces were analysed, which are shown in the section S9 of SI. It is observed that the PDMS surface is more electro-wettable than the PS surface. The applied voltage up to 100 V was not enough to bring a significant decrease in contact angle due to the pinning of TPCL of droplet on both PDMS and PS surfaces. Only beyond this voltage, the system gets enough energy to overcome the pinning effect, and the angle starts to decrease.

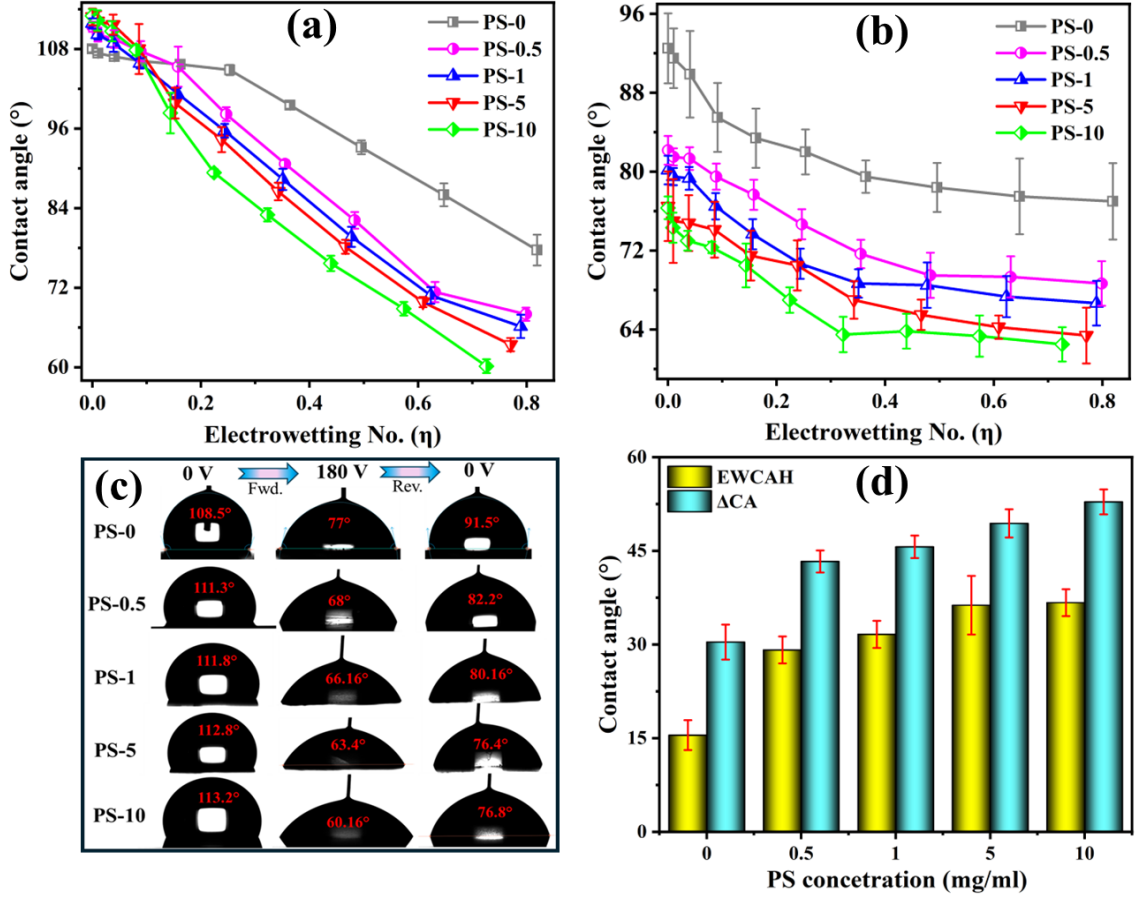

*Figure 4*: Electrowetting behaviour of PDMS surfaces with PS micro humps distributed on the surface. PS concentration was varied from 0 to 10 mg/ml. (a) Contact angle with increasing applied voltage from 0 to 180 volt and with (b) decreasing voltage from 180 volt to 0 volt. (c) Images of water droplets at different applied voltages for different samples. (d) Electrowetting contact angle hysteresis (EWCAH) and the maximum change in contact angle (*ΔCA*) under forward bias.

As reported earlier, being soft in nature, there is a formation of wetting ridge around the TPCL on the surface, causing easy pinning on the surface [31,32]. In the present study, for PDMS, the height ($h$) of ridge was estimated to be in the range of 45-90 nm using the relation, $h = \Upsilon_{la}/E$ where $E$ is the elasticity with a value of 0.8-1.6 MPa and $\Upsilon_{la}$ is the surface tension of water with a value of 72 mN/m [33,34]. Along with this ridge, the microscopic roughness as described in section S4 in SI also contributes to this pinning. Because of this pinning, the experimental curve of decreasing contact angle with applied voltage deviates considerably from the curve predicted by the L-Y equation. As detailed in section S10 of SI, the angle drops faster in the prediction compared to the experimental observation (Fig. S10 (a)). Even the modified L-Y equation, considering the roughness factor as mentioned in Eq. (2) failed to provide a fit to the data (Fig. S10 (a, b)). Therefore, L-Y equation requires further modification to follow the electrowetting behaviour of a real surface.

In Fig.4 (a) (increasing voltage), and (b) (decreasing voltage), the electrowetting behaviour measuring the water contact angle of the PDMS surface with a distributed PS humps on it is plotted as a function of the electrowetting number ($\eta$). The effect of the field differs from sample to sample with one of the reasons being the difference in specific capacitance of these samples as explained in Eq. (7) with varying '$a$' and '$h$' of the humps. In forward electrowetting, with the rise in PS concentration resulting in larger humps occupying a higher surface area, there is a downward shift of the curve. Thus, the presence of the micro-humps makes the surface more electro-wettable. In case of reverse electrowetting, on decreasing $\eta$, initially, there is a slow increase in contact angle which becomes rapid at a later stage. It is clear from the figure that after completing the electrowetting cycle, a droplet does not achieve its initial contact angle, which is also depicted in (Fig.4(c)). The difference of initial contact angle during forward electrowetting and the final contact angle on reverse electrowetting is termed as electrowetting contact angle hysteresis (EWCAH) [35], which increases with increasing the PS concentration. For a particular surface, the total change in contact angle ($\Delta$CA) from its initial to final value under the applied electric field during forward electrowetting is calculated and shown in Fig. 4 (d). The systematic increase of this value with PS concentration indicates the PS humps to facilitate electrowetting. While the presence of more PS humps (Fig. S11) increases the surface roughness slightly (Fig. 2(b)), it considerably boosts the surface energy (Fig. S5), which contributes to this higher electrowetting nature of the surface. Therefore, the chemical heterogeneity of a surface is a sensitive parameter to establish the nature of electrowetting of the surface. This heterogeneity is one of the deciding factors of a droplet pinning on a surface.

It is interesting to observe that the number of PS humps on the PDMS surface under a TPCL remains the same with an average number of 27 for a TPCL with a perimeter of 200 μm, even though the concentration of PS is increased (Fig. 5(a, b)). However, the average size of the humps monotonically increases with a value of 6 μm at the highest concentration of 10 mg/ml (Fig. 5(c)). Thus, the increasing size leading to a higher surface area of PS-micro-humps, and hence the higher surface energy, facilitates easy spreading of the water droplet during forward electrowetting. According to L-Y equation, for a surface to show high electrowetting nature, it must store high capacitive energy, which, from Eq. 1, is $\frac{\varepsilon_0 K V^2}{2d}$, in case the dielectric material is in a layer form. To achieve this higher energy, the value of dielectric constant ($K$) must increase while the overall thickness ($d$) must decrease with the introduction of PS humps. In the present study, the PS has nearly the same dielectric constant of ~2.5 as of PDMS. Again, with increasing PS concentration, the effective thickness of the layer increases. Therefore, the capacitive energy is solely not sufficient to explain the anomalous behaviour of the surface. Upon close examination, it is evident that the L-Y equation contains solely a dielectric capacitive term. It does not incorporate physicochemical heterogeneity. Further, it does not consider material elasticity, which controls the ridge height along the TPCL. All these surface characteristics decide a specific wetting nature of a surface, leading to deviation of experimental data from the L-Y predicted curves. Here, the L-Y equation can be modified by subtracting a surface parameter term ($P$) that includes the effects of surface morphology, its roughness, chemical heterogeneity and elasticity, providing an equation in the form,

$$cos\theta_E = cos\theta_0 + \left(\frac{1}{2\gamma_{lg}}\left(\frac{C}{A}\right) - P\right)V^2 \qquad (9)$$

In Fig. 6(a), the experimental data is successfully fitted by using this modified L-Y equation to extract the parameter $P$. Here, the calculated values of specific capacitance were used as

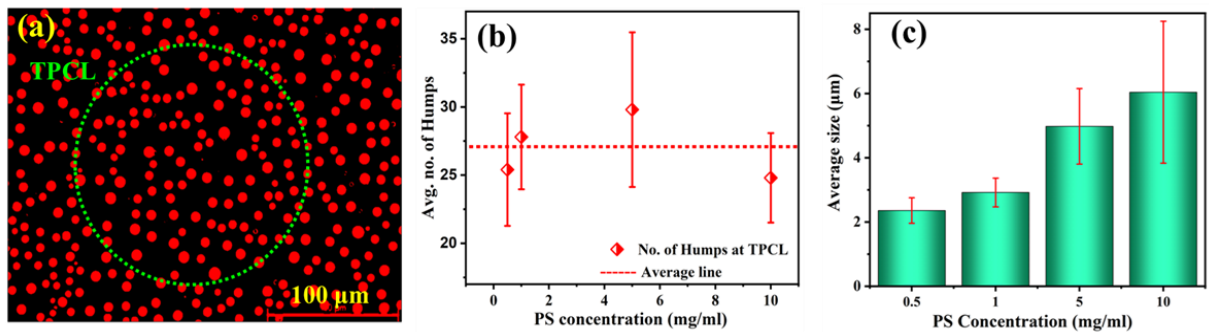

*Figure 5: (a) demonstration of triple phase contact line (TPCL) on PS micro-hump coated PDMS surface. (b) Average number of PS- humps underlying a TPCL. (c) Average size of PS-hump with increasing PS concentration.*

discussed in the previous section. For the pure PDMS surface, $P$ was found to be maximum of 12 μF/m-N, indicating maximum pinning. As the PS micro hump is introduced on PDMS surface, it starts to decrease. At 0.5 mg/ml of PS, $P$ decreases to 0.5 μF/m-N, after which, it shifts to negative values (Fig. 6(b)). The positive value of surface parameter represents the resistive nature of the surface in spreading the TPCL, while the negative value supports the spreading. $P = 0$ indicates the surface to behave ideally, which follows the L-Y equation. As shown in the inset of the Fig. 6(b), the threshold voltage, which initiates spreading of a droplet under applied voltage, decreases with increasing PS concentration [36]. This is a direct reflection

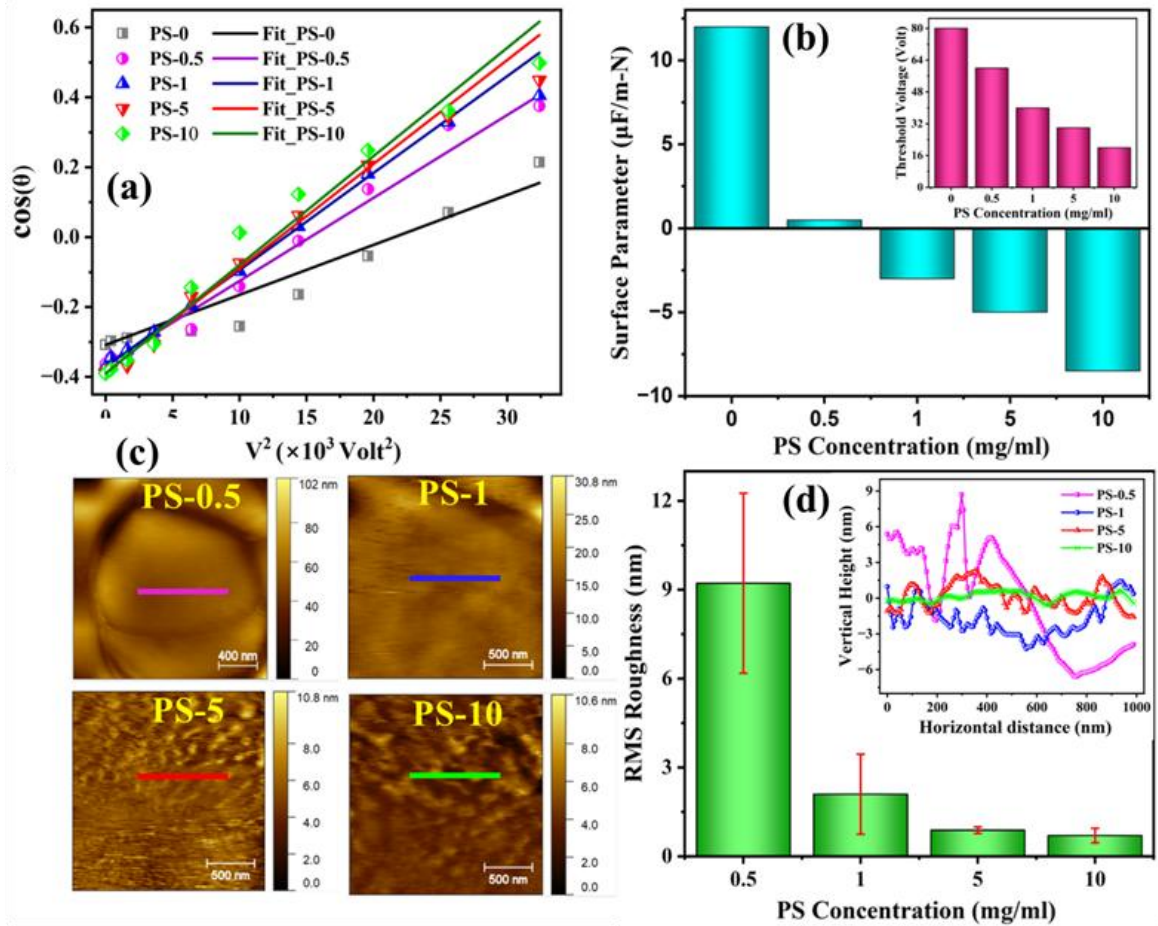

*Figure 6:* (a) Fitting to experimental data with a modified Lippman-Young equation discussed in the text (Eqn. 9). Symbols show the experimental data with the solid line being the best fit. (b) Extracted surface parameter from the fit with increasing PS concentration. Inset shows the threshold voltage of modified surface. (c) AFM images of PS micro-humps surface with the measured RMS roughness shown in (d). The inset of (d) shows the line profile of PS micro-hump surfaces exhibiting lower roughness at higher concentration of the PS producing larger hump.

of weaker pinning of the water droplet on the PS-hump distributed PDMS surface. This weaker pinning and hence the drop in the $P$ parameter is the consequence of multiple factors as predicted above. It is evident in Fig. 6(c) that a larger hump lowers the roughness of the PS occupied surface. Firstly, for the PS concentration of 0.5 mg/ml, the micro-humps have the maximum roughness of ~9.2 nm which drops to 0.7 nm at the concentration of 10 mg/ml. The line profile of the PS surface unambiguously exhibits this conclusion (Fig. 6(d)). A PS surface becomes weakly hydrophilic on reducing its surface roughness. It helps spreading the water droplet on the micro-humped surface. Secondly, enhancing the surface area of a high surface energy material over a lower one helps spreading water. More surface of PS reduces the Gibbs barrier that define the energy cost for moving the TPCL over chemical heterogeneities between

PS and PDMS. The mismatch between the surface energy of PS and PDMS has been discussed in the earlier section. Thirdly, the PS surface is harder (~3.3 GPa [37]) than the PDMS (~1MPa [38]) surface that leads to a lower ridge around the TPCL.

The physicochemically heterogeneous surface in the present study with the high surface energy PS micro-humps distributed on a low surface energy PDMS film is an important engineered surface. Such dual-material surfaces appear in several advanced technologies. PS micro-humps on PDMS creates strong chemical and topographical contrast, which can be useful for tuneable wetting and liquid manipulation on surfaces. PS is biocompatible which supports cell adhesion while the PDMS is usually cell-repellent. Therefore, this patterned surface can guide cell alignment, confinement and cell patterning for diagnostics. PS humps provide hard points on PDMS soft matrix, creating hybrid mechanical behaviour which is applicable for micro-tribology test platforms, controlled friction pads, and soft robotics skins with tuned friction. This surface can also be useful for a strain and pressure sensor as it has the spots of hard PS humps which is difficult to deform and the soft elastic PDMS, which is easy to deform.

Along with such wide applications these surfaces have also some limitations. Because of the energy mismatch between PS humps and the flat PDMS layer, the TPCL experiences a local energy barrier, causing pinning/depinning with a stick-slip nature during electrowetting. Further, the surface exhibits a non-uniform electric field distribution with the application of an external field. Note that PS is a strong electrical insulator and susceptible to charge trapping. Consequently, charges accumulate around micro-humps, resulting in a varied responsiveness to the applied field.

From a future perspective, a systematic study could be conducted in which hump height is varied independently of diameter, spacing, and shape (e.g., cylindrical pillars, conical structures) to identify a geometry that minimize contact angle hysteresis and maximize droplet controllability. To eliminate pinning and enhance droplet mobility under electrowetting, a study involving the introduction of a thin lubricating fluid trapped between the PS humps could be investigated. This system could further be integrated into a multi-responsive droplet platform, in which droplets containing magnetic nanoparticles can be controlled by more than one external stimulus. Thus, these future investigations will make such a surface to enable smooth, reversible, and low-stimuli droplet manipulation for microfluidic and sensing applications.

# 4. Conclusion

The present study investigates the electrowetting behaviour of a polydimethylsiloxane (PDMS) surface coated with polystyrene (PS) micro-humps. The size, shape and roughness of these humps were characterized by optical microscopy and atomic force microscopy and FESEM images. These micro humps were found identical in shape, while their sizes follow a Gaussian distribution. The electrowetting nature of the surface was enhanced with increasing PS concentrations. The results were found to deviate from the Lippmann-Young (L-Y) equation. The presence of micro-humps, which become less rough with increasing their size, raises the effective surface energy and roughness, thereby a surface leading to decrease in Gibbs barrier during applied voltage. It makes the surface more electro-wettable. To extract the effects induced due to surface modification, a surface parameter, $P$ was introduced into the electrowetting equation. The positive value of $P$ corresponds to contact line pinning, whereas negative value indicates depinning/spreading of the droplet on surface. The calculated values of surface parameter confirm faster electrowetting than predicted by the classical L-Y equation. Furthermore, the electrowetting contact angle hysteresis increases with PS concentration. Overall, this present study opens a way of understanding and provide new insight into electrowetting nature on a nano- or micro-structured heterogeneous surface.

# 5. Acknowledgement

S. K. G. acknowledges the Board of Research in Nuclear Sciences (BRNS), Govt. of India for the financial support to procure the instrument 'Optical Tensiometer' under project no. 58/14/13/2022-BRNS/ 37059. All the authors thank Dr. Ravindra Singh and Dr. Raju Vemoori for their technical support in using AFM and FESEM, respectively.

# Supporting Information (S.I.)

**Anomalous electrowetting of physicochemically heterogeneous surfaces**

Rumal Singh, Donjo George, Prashant Hitaishi[#], Samarendra P Singh, Sajal K Ghosh[*]

*Department of Physics, School of Natural Sciences, Shiv Nadar Institution of Eminence,*
*NH-91, Tehsil Dadri, Gautam Buddha Nagar, Uttar Pradesh-201314, India*
*\*Email: sajal.ghosh@snu.edu.in*

[#]*Present address: Institute of Experimental and Applied Physics, Kiel University, Germany*

## S1. Electrowetting experiment setup

The electrowetting circuit was completed by connecting one end of the wire to an ITO glass substrate and another end to a platinum wire that was dipped in water droplet placed on the PDMS-coated ITO glass substrate, as illustrated in Figure S1. As shown in the figure, $\theta_o$ is the initial contact angle, which changes to $\theta_E$ on applied external voltage which spreads the liquid on the surface, hence reducing the contact angle.

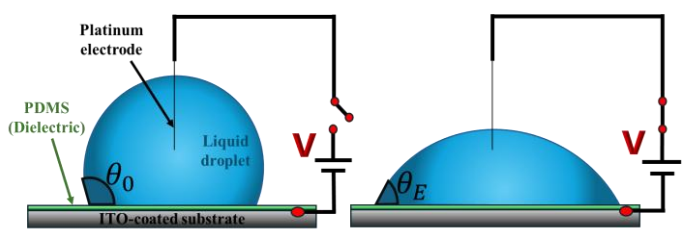

***Figure S1:*** *Schematic diagram of electrowetting experiment with no applied voltage (V=0) and a finite applied voltage (V $\neq$ 0).*

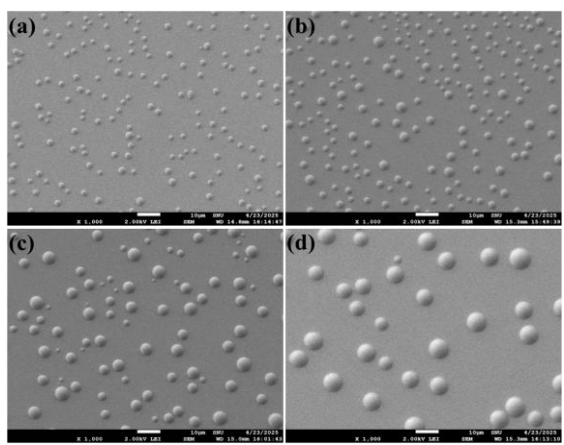

***Figure S2:*** *FESEM images of polystyrene (PS) micro-humps on polydimethylsiloxane (PDMS) surface. (a-d) top views of the humps at a PS-concentration of 0.5, 1.0, 5.0, 10.0 mg/ml respectively.*

## S2: FESEM images of micro-humps and their size distribution

The top view of FESEM images of the micro-humps are shown in Figure S2. It is evident that the size of the micro humps depends on the polystyrene (PS) concentration in the chloroform solution. The shape of micro humps remains same for all concentration while size increases with increasing PS concentration. This size distribution is fitted with Gaussian distribution function expressed as,

$$F(x, \mu, \sigma) = \frac{1}{\sigma\sqrt{2\pi}} [\exp\left\{\frac{(x-\mu)^2}{2\sigma^2}\right\}] \qquad (S1)$$

where $x$ is the number of micro humps, $\mu$ defines the centre of distribution and $\sigma$ is the standard deviation that controls the dispersion of distribution. For micro-humps formed with varying PS concentration on PDMS, the values of parameters $\mu$ and $\sigma$ are shown in Table S1. The increase in value of $\mu$, indicates that the characteristic hump size grows with PS concentration, while the increase in value of $\sigma$ shows that the size distribution becomes broader (more dispersed).

***Table S1:*** *The centre of distribution ($\mu$) and the standard deviation ($\sigma$) of the PS micro-humps on PDMS surface with increasing PS concentration.*

| PS Concentration (mg/ml) | $\mu$ (μm) | $\sigma$ (μm) |
|---|---|---|
| 0.5 | 2.357 | 0.395 |
| 1 | 2.918 | 0.447 |
| 5 | 4.978 | 1.175 |
| 10 | 5.985 | 2.041 |

## S3. Wetting parameters of polystyrene (PS) and polydimethylsiloxane (PDMS)

Figure S3 shows the advancing (Adv.) and receding (Rec.) contact angle (CA) of a water droplet on a pure PS and PDMS surfaces. While PS shows an Adv. CA of 97.40 ± 3.99°, the PDMS exhibits it to be 117.00 ± 0.46°. A similar trend is observed for Rec. CA. Therefore, it is conclusive that the PDMS surface is more hydrophobic compared to a PS surface.

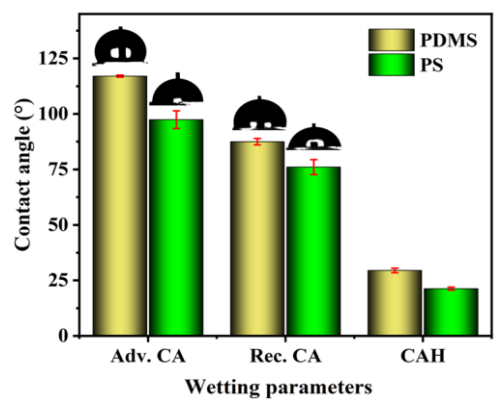

***Figure S3:*** *Wetting behaviour of polystyrene (PS) and polydimethylsiloxane (PDMS) surfaces showing the advancing contact angle (Adv. CA), receding contact angles (Rec. CA) and the contact angle hysteresis (CAH) of a water droplet.*

## S4. AFM images of PDMS and PS surfaces

The roughness of the PDMS and the PS surfaces are quantified by AFM images, as shown in Fig. S4. It is evident that the PDMS ($R_{rms} = 0.53\ nm$) surface is rougher than the PS ($R_{rms} = 0.20\ nm$) one.

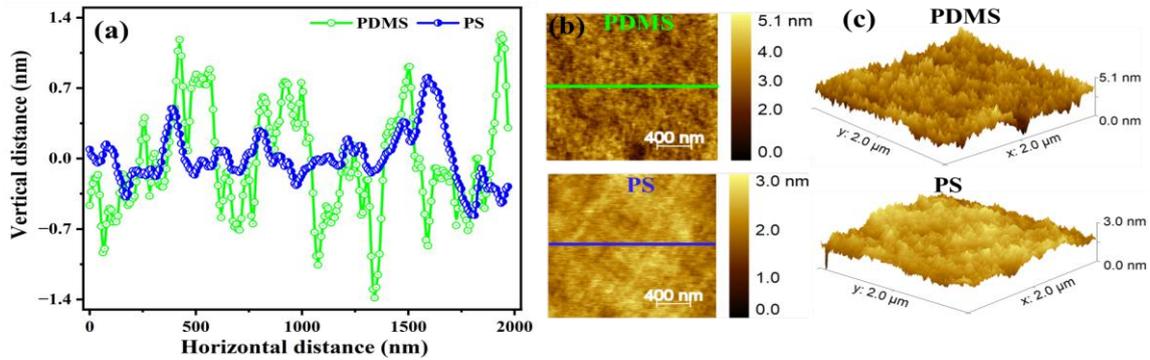

***Figure S4:*** *(a) The height profile of the line cut of the AFM images of the PDMS and PS surfaces. (b) Top view of the respective surfaces and (c) the three-dimensional view of them. Data exhibit that the PDMS surface is rougher than the PS one.*

## S5. Net surface energy

The PDMS surface energy (19-21 mJ/m$^2$) is almost half of PS (40-44 mJ/m$^2$) surface energy [1]. The PS micro-humps on the PDMS surface are formed in isolation from one another. The net surface energy of the surfaces is computed as the sum of their individual surface energies multiplied by their respective area fractions. Equation S2 obtained the net value of the energy which shows that with increasing PS concentration the surface energy increases (Fig. S5).

$$S_{net.} = f\ S_{PS} + (1-f)S_{PDMS} \tag{S2}$$

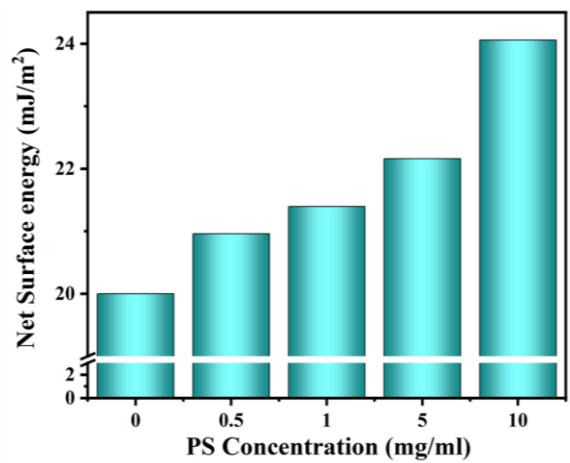

***Figure S5:*** *Change in net surface energy with increasing PS concentration on the PDMS thin layer on top of ITO coated glass surface.*

## S6. Calculation of roughness factor of substrate

The roughness factor of PS coated PDMS surface was calculated using the optical and FESEM images. The optical images were captured in the size of $300\ (L) \times 225\ (W)\ \mu m^2$. These images were processed with Image-J software, where the microdots were marked with a circle of radius R. FESEM image of cross section exhibits them as spherical cap with a contact angle ($\theta$) of ~ 45°. These structures are referred as a 'micro-hump' as shown in Fig. S6. From the optical and FESEM images, the interfacial area of micro-hump exposed to air was calculated from the expression, $A_I = 2\pi R^2(1 - \cos\theta)/\sin^2\theta$. The contact interfacial area of the PS droplet with PDMS surface was obtained from, $A_c = \pi R^2$. The roughness factor (R) was then determined by $R = \frac{\text{Actual area }(A_a)}{\text{Projected area }(A_p)}$. The actual area ($A_a$) and the projected area ($A_p$) of substrate were calculated with the following expressions,

$$A_p = L \times W \tag{4}$$

$$A_a = A_p + \sum_{n=1}^{n} A_I - \sum_{n=1}^{n} A_c \tag{5}$$

where the *n* is the total number of micro-humps present in captured image of length L and width W.

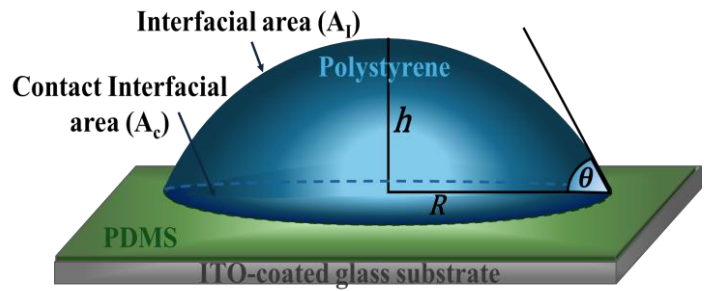

*Figure S6:* Schematic diagram of a polystyrene (PS) micro-hump on polydimethylsiloxane (PDMS) surface showing its base radius R and height h.

## S7. PDMS film thickness measurement

To measure the PDMS film thickness coated on ITO-coated substrate, it was peel-off from a small region with a fine cut. After that the AFM image (Fig.S7) was taken with a region consisting of half part with PDMS film and half without PDMS as shown in Fig.S7. Thus, the height between the two points, one at peel-off region and other at PDMS surface was measured which is the thickness of PDMS film. The thickness measurement was done at 10 different regions over the surface, and the statistically averaged value is found to be 4.92 ± 0.54 µm.

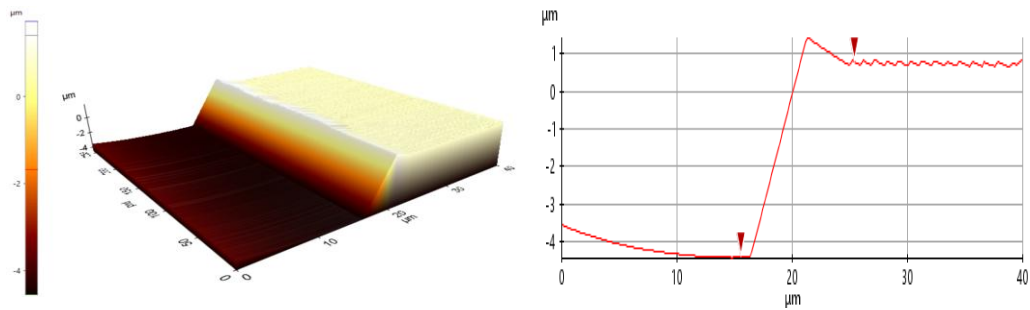

*Figure S7:* The thickness measurement of Pure PDMS film by atomic force microscopy (AFM).

## S8. Effective specific capacitance

In the present system, (Fig. S8), there are two different capacitors, one is simple PDMS layer of $d$ thickness, and the other is composite system where the same PDMS layer of thickness $d$, having PS micro-hump. Thus, the effective capacitance of complete system becomes the parallel combination of both, pure PDMS layered capacitance $\left(C_l = \frac{K\epsilon_0 A}{d}\right)$ and the composite system ($C_c$). Since the dielectric constant of PDMS and PS are very close to each other, it is considered to be same which is denoted by $K$.

To get the specific capacitance, of composite system, the ITO-coated surface where PDMS is coated, is considered z-plane as shown in Fig.S8. The point 'O' in z-plane where the axis symmetry of PS-hump intersects perpendicularly, is considered as origin O (0,0,0). The hump (green solid line), which is a spherical section of a sphere (yellow dotted line) of radius '$R$' is centred at O' $(0,0, d - R \cos \theta)$, with height '$h$', base contact radius '$a$', and contact angle '$\theta$' is situated on top of PDMS at point C $(0, 0, d)$.

From the geometry, in Fig. S8,

$$a = R \sin \theta \qquad (S1)$$
$$h = R(1 - \cos \theta) \qquad (S2)$$

*Local separation between z-plane ($z = 0$) and point P (x, y, z) at hump's curved surface:* In this model, the local distance between the z-plane ($z = 0$, ITO-coated surface) and point P (x, y, z) located at hump's curved surface is not constant. It varies from point to point due to the curved surface. Therefore, first, the local distance $z(r)$, is calculated in terms of $d$, $h$, $R$ and $r = \sqrt{x^2 + y^2}$. From the equation of sphere of radius R, centred at $(0, 0, d - R\cos\theta)$,

$$(x-0)^2 + (y-0)^2 + \{z - (d - R\cos\theta)\}^2 = R^2$$

$$x^2 + y^2 + \{z - (d - R\cos\theta)\}^2 = R^2$$

$$\{z - (d - R\cos\theta)\}^2 = R^2 - (x^2 + y^2) \tag{S3}$$

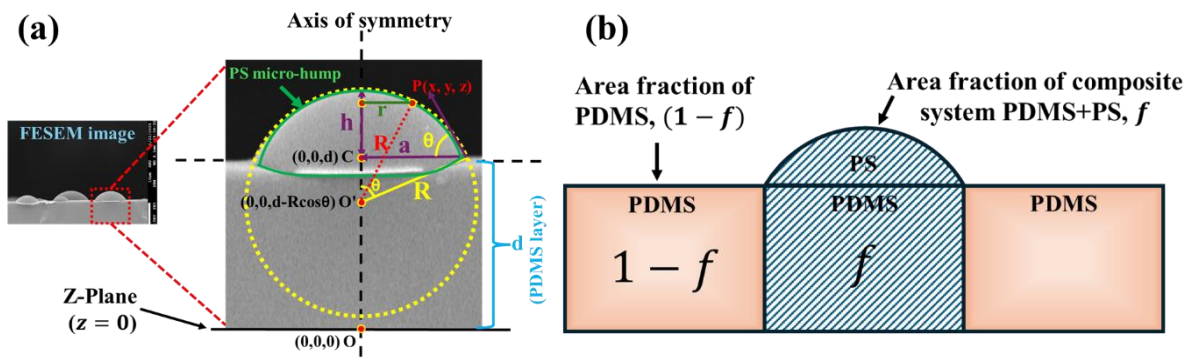

*Figure S8:* *The schematic representation of (a) PS coated PDMS surface for calculating the specific capacitance, and (b) area fraction of PDMS and composites system to calculate the effective capacitance of complete system.*

Let, $(x^2 + y^2) = r^2$, where r is the distance of a point P from axis of symmetry, situated at the curved surface.

$$\{z - (d - R\cos\theta)\}^2 = R^2 - r^2$$

$$z - (d - R\cos\theta) = \pm\sqrt{R^2 - r^2} \tag{S4}$$

For upper positive half of sphere,

$$z - (d - R\cos\theta) = \sqrt{R^2 - r^2}$$

$$z = d - R\cos\theta + R\left(1 - \frac{r^2}{R^2}\right)^{\frac{1}{2}} \tag{S5}$$

On expanding binomially, the last term on the right side and neglecting the higher order terms in the expansion,

$$z = d - R\cos\theta + R\left(1 - \frac{r^2}{2R^2}\right)$$

$$z = d - R\cos\theta + R - \frac{r^2}{2R}$$

$$z = d + R(1 - \cos\theta) - \frac{r^2}{2R} \tag{S6}$$

Using, equation S2,

$$z = d + h - \frac{r^2}{2R} \tag{S7}$$

This equation (S7) gives the local separation z, at each point $d(x, y)$ between the flat and curved surface.

*Capacitance of composite (PDMS+PS) system, $(C_c)$:* To calculate capacitance of the composite system, as shown by shaded region in Fig. S8(b), system with flat surface area $A = \pi r^2$ can be calculated by integrating the differential capacitance element from 0 (axis of symmetry) to $a$ which is the base contact radius of hump. Differential surface element $dA = 2\pi r dr$. The differential capacitance element, $dC = \frac{\varepsilon_0 dA}{z(r)} = \frac{2\pi\varepsilon_0 r dr}{d + h - \frac{r^2}{2R}}$. Now, the capacitance of composite system can be achieved by integrating $dC$ over a limit from 0 to $a$.

$$C = \int_0^a dC$$

$$C = \int_0^a \frac{2\pi K\varepsilon_0 r}{d + h - \frac{r^2}{2R}} dr \quad (S8)$$

Let, $d + h - \frac{r^2}{2R} = u$, then $dr = -\frac{R}{\sqrt{2R\{(d+h)-u\}}} du$ and the limit will be changed.

For, $r = 0, \Rightarrow u = d + h$ and for $r = a, \Rightarrow u = d + h - \frac{a^2}{2R}$

$$C = -\int_{d+h}^{d+h-a^2/2R} \frac{2\pi K\varepsilon_0 \sqrt{2R\{(d+h)-u\}}}{u} \frac{R}{\sqrt{2R\{(d+h)-u\}}} du$$

$$C = -2\pi K\varepsilon_0 R \int_{d+h}^{d+h-a^2/2R} \frac{1}{u} du$$

$$C = -2\pi K\varepsilon_0 R (\ln u)_{d+h}^{d+h-a^2/2R}$$

$$C = -2\pi K\varepsilon_0 R \left\{ \ln\left(d + h - \frac{a^2}{2R}\right) - \ln(d+h) \right\}$$

$$C = -2\pi K\varepsilon_0 R \ln\left\{1 - \frac{a^2}{2R(d+h)}\right\} \quad (S9)$$

From equation (S1), $R = \frac{a}{\sin\theta}$, inserting in above equation,

$$C = -2\pi K\varepsilon_0 \frac{a}{\sin\theta} \ln\left\{1 - \frac{a\sin\theta}{2(d+h)}\right\} \quad (S10)$$

*Specific Capacitance ($C_s$):* The specific capacitance can be calculated by dividing the capacitance by area equal to the contact area of PS-hump with PDMS surface as follows,

$$C_s = \frac{C_c}{A}$$

$$C_s = -\frac{2\varepsilon_0}{a\sin\theta} \ln\left\{1 - \frac{a\sin\theta}{2(d+h)}\right\} \quad (S11)$$

*Effective specific capacitance:* To calculate the effective specific capacitance, the effect of pure PDMS layered ($C_l$) should be added where the PS micro-hump is absent. This can be realised by considering a parallel combination of two capacitors; one is a parallel plate due to pure PDMS layer ($C_l$), and another is a flat-curved plate capacitor formed by composite system ($C_c$), multiplied by their surface area fraction (Fig. S7(b)) as follows,

$$C_{Eff.} = f.C_c + (1-f).C_l \quad \textbf{(S12)}$$

## S9. Electrowetting behaviour of pristine PDMS and Polystyrene surfaces

The electrowetting behaviour of PDMS and PS surfaces are shown in Fig. S9. It shows contact angles when electric potential is varied under forward (Fwd.) and reverse (Rev.) bias in a range of 0 to 210 V. The maximum voltage was limited to 210 V to avoid the breakdown of the dielectric film. The initial contact angle for PDMS is 108 ± 1° while for polystyrene is 88 ± 1° (Fig. S9(c)) reflecting PDMS to be more hydrophobic than the polystyrene. The high hydrophobicity of PDMS can be attributed to lower surface energy (~ 19-21 mJ/m²) compared to PS (~ 40-44 mJ/m²) [1]. In the forward (Fwd.) electrowetting, when the applied voltage is increased up to ~ 100 V on both surfaces, there is only slight decrease in contact angle (Fig. S9(a, b)). Upon further increasing the applied voltage, the contact angle decreases considerably. In case of PDMS, it reaches to a minimum of 63 ± 1° and for PS it is 61 ± 2° (Fig. S9(c)). In reverse (Rev.) electrowetting, it is observed that the contact angle does not follow the same path as of Fwd. electrowetting. In Fwd. electrowetting, the contact angle initially decreases slowly with a rate of ~0.03 °/V. After 100 V, the rate increases to 0.35 °/V for PDMS. For PS, the respective rates change from 0.024 °/V to 0.27 °/V which is slower than PDMS. At the end of Fwd. electrowetting, the total drop in contact angles is 45°, and 25° for PDMS and PS, respectively. Thus, the PDMS surface is more electro-wettable than the PS surface. At the end of Rev. electrowetting the achieved contact angle is 93° for PDMS and 75.5° for PS. The difference between the initial contact angle in Fwd. electrowetting and the contact angle at the end of Rev. electrowetting is termed as electrowetting contact angle hysteresis (EWCAH). It reflects the energy loss during droplet spreading and retraction. The values of EWCAH for PDMS and PS are 15° and 10°, respectively. Therefore, the energy loss for PDMS is more than the PS.

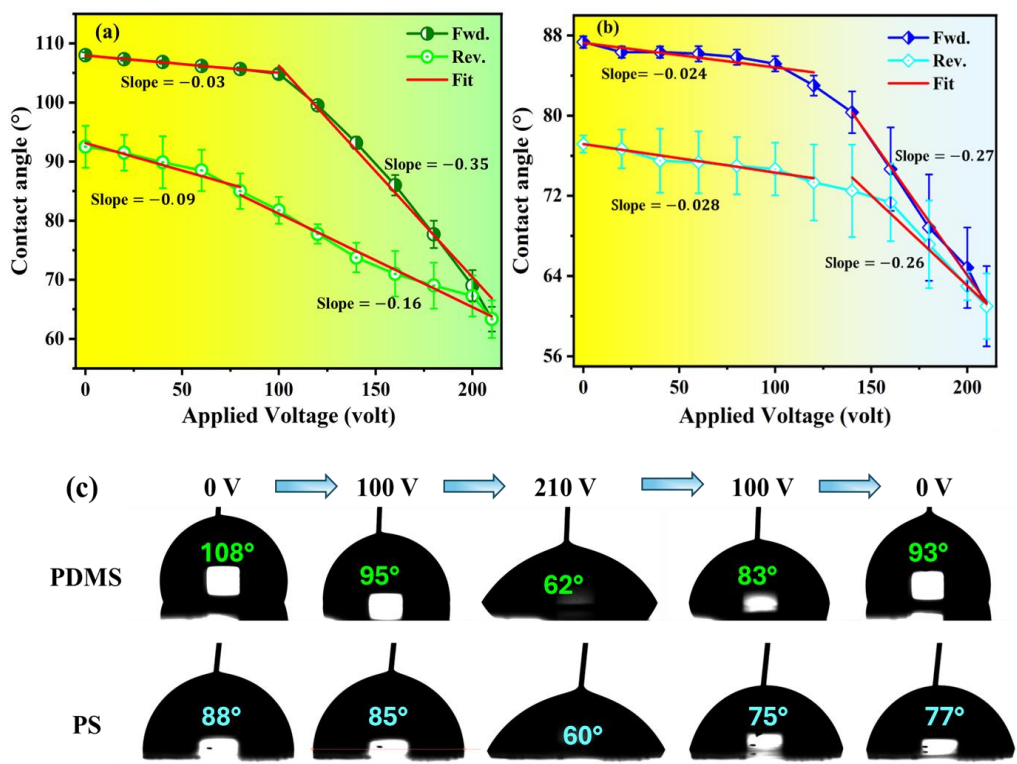

**Figure S9**: *Electrowetting behaviour of (a) PDMS and (b) PS coated substrate in forward (Fwd.) and reverse (Rev.) applied voltage. (c) captured images of electro-wetted surface at varying voltage in a cycle.*

When a droplet is deposited on a sufficiently soft, it can deform the surface and form a wetting ridge around the triple phase contact line (TPCL) [2]. This line is then pinned into it. Also, the microscopic roughness of the surface contributes to this pinning. In the present system, the applied voltage up to 100 V is not enough to overcome this pinning. Only beyond 100 V, the system gets enough energy, leading to a decrease in contact angle.

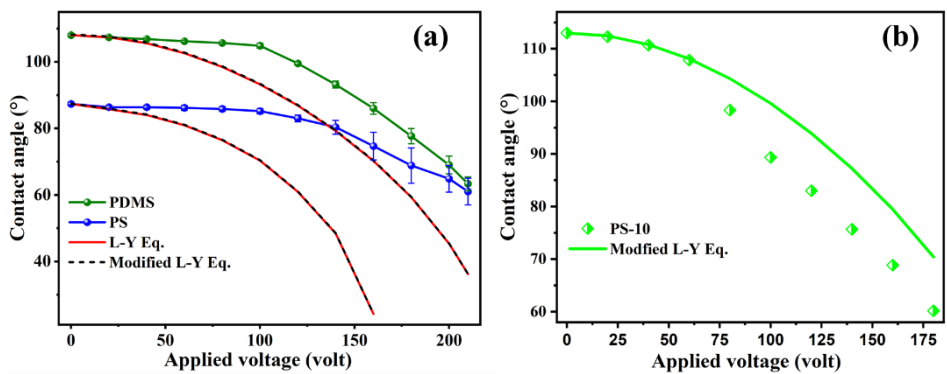

*Figure S10:* *(a) Deviation of theoretical prediction using Lipmann-Young (L-Y) equation (solid redline and dotted line) from experimental data (line and symbol) of PDMS and PS surfaces. (b) Prediction using modified L-Y equation for PS micro humps decorated PDMS surface (solid green line) and the data (symbol).*

## S10. Deviation of electrowetting nature from theoretical curve

A comparison of electrowetting behaviour of PDMS and PS surfaces is shown in Fig. S10 (a). The deviation of experimental results from the Lippmann-Young (L-Y) equation (solid red lines) was observed. PDMS is softer than PS [3] which causes easy pinning on PDMS, favouring higher EWCAH. The surface roughness and material impurities also contribute to this lag between the curve of decreasing contact angle with applied voltage obtained from the L-Y equation and experiment (Fig S10). The EWCAH, which is 15° for PDMS and 10° for PS, is found to exhibit a similar trend to contact angle hysteresis (CAH) of ~29.5° (PDMS) and ~21° (PS). The CAH reflects purely surface properties like roughness and its chemistry, while EWCAH combines electrical modulation with it. To include the effect of surface roughness, the L–Y equation was modified to $cos\theta_E = R(\cos\theta_0 + \eta)$ as shown in equation (2) of the main manuscript. This modified equation was used to follow the data of pure PDMS and PS surface but without much success (dotted line in Fig. S10(a)). The dotted and the solid lines are very close to each other as the surfaces have a very low roughness.

The electrowetting nature of PS micro hump decorated PDMS surface is shown in Fig. S10 (b) for the sample with 10 mg/ml PS in chloroform. The solid line shows the modified L-Y equation as mentioned above, considering the roughness of the composite surface (solid line), but the theoretical line deviates considerably from the experimental data. Therefore, only the roughness of the surface cannot explain the electrowetting behaviour of a real sample.

**S11. Morphology and roughness factor from AFM images**

The AFM study was performed on a surface of size $5 \times 5$ μm$^2$ in order to observe the morphology of the PS-coated PDMS surface. Figure S11(a) displays the images with increasing PS concentration, while Figure S11(b) displays the roughness factor that was computed from these images.

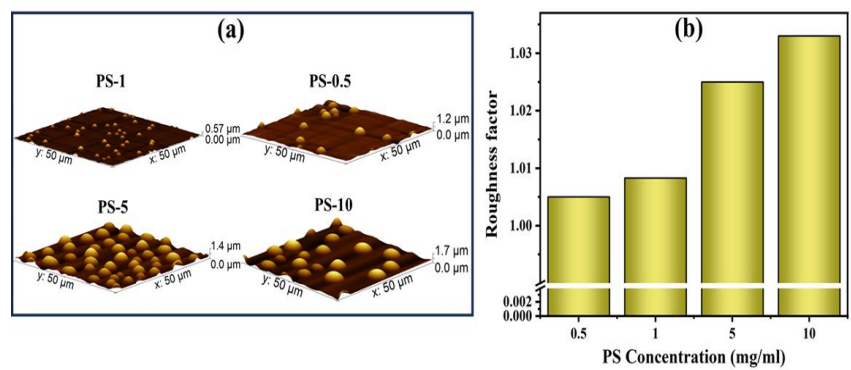

***Figure S11:*** *(a) AFM images of PS micro-hump coated PDMS surface. (b) Roughness of the surfaces with increasing PS concentration.*